\documentclass{iau} 
\usepackage{graphicx}
\usepackage{natbib}
\usepackage{aas_macros}
\title[Milky Way Nuclear Star Cluster] 
{Observational constraints on the formation and evolution of the Milky Way nuclear star cluster with Keck and Gemini}

\author[Do et al.] 
{Tuan Do$^1$, Andrea Ghez$^1$, Mark Morris$^1$, Jessica Lu$^2$, Samantha Chappell$^1$,  Anja Feldmeier-Krause$^3$, Wolfgang Kerzendorf$^3$, Gregory David Martinez$^1$, Norm Murray$^4$, \and Nathan Winsor$^5$}

\affiliation{$^1$Department of Physics and Astronomy \\ 430 Portola Plaza, Box 951547 
Los Angeles, CA, 90095-1547  \\ email: {\tt tdo@astro.ucla.edu} \\[\affilskip]
$^2$Dept. of Astronomy, University of California, Berkeley Berkeley, CA, 94720, USA \\[\affilskip]
$^3$European Southern Observatory (ESO), Karl-Schwarzschild-Straße 2, 85748 Garching, Germany\\[\affilskip]
$^4$Canadian Institute for Theoretical Astrophysics, University of Toronto, 60 St. George Street, Toronto, ON M5S 3H8, Canada\\[\affilskip]
$^5$Department of Astronomy \& Astrophysics, University of Toronto, 50 St. George Street, Toronto M5S 3H4, ON, Canada}

\pubyear{2016}
\volume{322}  
\jname{The Multi-Messenger Astrophysics of the Galactic Centre}
\editors{Roland Crocker, Steve Longmore, \& Geoff Bicknell, eds.}
\begin{document}

\maketitle

\begin{abstract}

Due to its proximity, the Milky Way nuclear star cluster provides us with a wealth of data not available in other galactic nuclei. In particular, with adaptive optics, we can observe the detailed properties of individual stars, which can offer insight into the origin and evolution of the cluster. We summarize work on the central parsec of the Galactic center based on imaging and spectroscopic observations at the Keck and Gemini telescopes. These observations include stellar positions in two dimension and the velocity in three dimensions. Spectroscopic observations also enable measurements of the physical properties of individual stars, such as the spectral type and in some cases the effective temperature, metallicity, and surface gravity. We present a review of our latest measurements of the density profiles and luminosity functions of the young and old stars in this region. These observations show a complex stellar population with a young (4-6 Myr) compact star cluster in the central 0.5 pc embedded in an older and much more massive nuclear star cluster. Surprisingly, the old late-type giants do not show a cusp profile as long been expected from theoretical work. The solution to the missing cusp problem may offer us insight into the dynamical evolution of the cluster. Finally, we also discuss recent work on the metallicity of stars in this region and how they might be used to trace their origin. The nuclear star cluster shows a large variation in metallicity ([M/Fe]). The majority of the stars have higher than solar metallicity, with about 6\% having [M/Fe] $< -0.5$. These observations indicate that the NSC was not built from the globular clusters that we see today. The formation of the nuclear star cluster is more likely from the inward migration of gas originating in the disk of the Milky Way

\end{abstract}

\firstsection 
\section{Introduction}

\begin{figure*}[tbh]
\center
\includegraphics[width=6in]{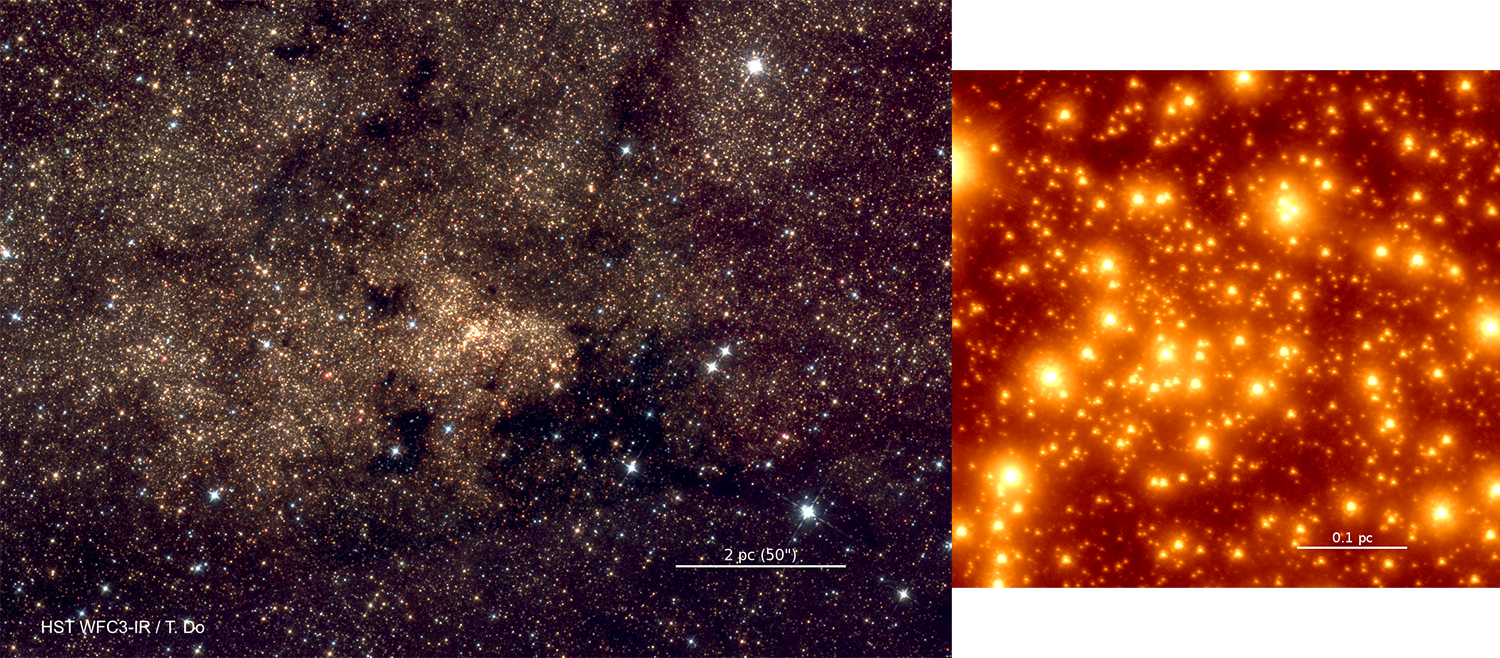}
\caption{Left: HST/WFC3-IR mosaic of the MW NSC (from GO-12182, PI: Do). Right: A laser-guide star adaptive optics image of the center of the cluster with the NIRC2 imager on Keck 2. The Galactic plane is oriented left to right in these images.}
\label{fig:nsc}
\end{figure*}

Since its discovery in the near-infrared \citep{1968ApJ...151..145B}, the Milky Way nuclear cluster (MW NSC) has provided us with insight as to the origin and evolution of the nucleus of Milky Way-like galaxies (Figure \ref{fig:nsc}). With the advent of imaging with adaptive optics, stars can be individually resolved down to tens of AU from the supermassive black hole that lies at the heart of the cluster \citep[e.g.][]{2007A&A...469..125S,2008ApJ...689.1044G,2009ApJ...692.1075G,2016arXiv160705726B}. Spectroscopy behind adaptive optics in particular has revolutionized our understanding of this region \citep[e.g.][]{2006ApJ...643.1011P,2009ApJ...703.1323D}. In this paper, we will primarily review the past decade of observations from the Keck and Gemini Telescopes in the central 1 pc of the MW NSC. Adaptive optics observations of this region has had the largest scientific impact because stellar crowding in this region has largely hidden this region from seeing-limited observations. We will discuss three main topics concerning the stellar components of the cluster: (1) the missing cusp problem, (2) the paradox of youth, and (3) the origin of the MW NSC. We also discuss the current limitations of the observations and offer suggestions as to the path forward.  


\section{The Missing Cusp Problem}

\begin{figure*}[tbh]
\center
\includegraphics[width=5.25in]{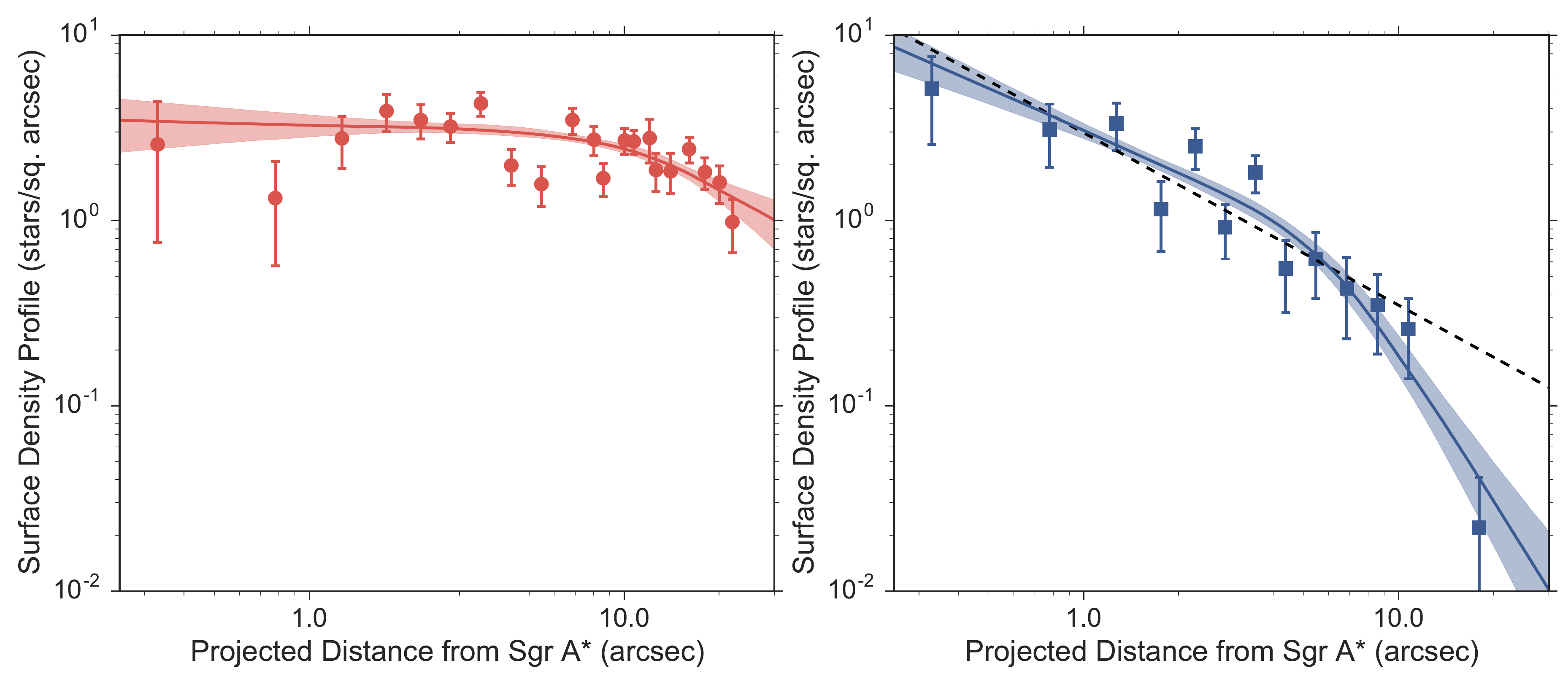}
\caption{Stellar surface density profiles as a function of projected distance from Sgr A* of late-type (left) and early-type (right) stars. The line shows the best fit radial profile for the late-type (old, $>1$ Gyr) and early-type (young, $4-6$ Myr) stars along with the 1 $\sigma$ shaded regions for the uncertainty in the model. 10 arcsec corresponds to 0.4 pc at a distance of 8 kpc to the Galactic center. The fit uses the Bayesian inference method described in \citet{2013ApJ...764..154D} to avoid potential biases from binning. To visualize the quality of the fit though, we plot radially binned surface density measurements from Keck and Gemini that have been corrected for completeness, as detailed by \citet{2013ApJ...764..154D} and \citet{2015ApJ...808..106S}. The late-type giants show a core profile in the inner $\sim 0.4$ pc. The dashed line shows the fit to the density profile of young stars only inside of $\sim10^{\prime\prime}$. The surface density of young stars drops rapidly beyond this radius, likely indicating that this is the edge of the cluster.}
\label{fig:density}
\end{figure*}

The missing cusp problem comes from the expectation that while there should be a steeply increasing stellar density profile toward the supermassive black hole, observations have shown a very flat density profile instead. A dense stellar cusp has been predicted from many theoretical works starting in the 1970s \citep{1976ApJ...209..214B,1977ApJ...216..883B}. Subsequent theoretical work has confirmed that star clusters with a massive central black hole should show a cusp if the cluster is dynamically relaxed \citep[e.g.][]{1991ApJ...370...60M,2009ApJ...697.1861A}. The Galactic center is the ideal place to look for such a cusp because it contains both a supermassive black hole that has been independently verified \citep[e.g.][]{2016arXiv160705726B} and an old stellar population with an age comparable to the timescale for two-body relaxation in this region. While early adaptive optics imaging showed a cusp in the stellar number density profile \citep{2003ApJ...594..812G}, it became clear with spectroscopy that the old late-type giants do not have a cusp \citep{2009ApJ...703.1323D,2010ApJ...708..834B}. The cusp seen in the number counts is from the large number of early-type stars, which are too young (see Section \ref{sec:young}) to be dynamically relaxed. Using the OSIRIS integral-field spectrograph at Keck, we found that the number density of early-type stars dominates in the near-infrared inside of the central 3 arcseconds (0.1 pc) from Sgr A*. The late-type giants on the other hand, have a core-like profile in this region. This density profile has been extended out to 1 pc with observations from the NIFS AO instrument on Gemini North \citep{2015ApJ...808..106S}. We combined these observations together to produce the deepest spectroscopic determination of the late-type and early-type stellar density profiles so far (Figure \ref{fig:density}). 

We use a broken power law model to fit the surface density profile $\Sigma(R)$ as a function of projected distance $R = \sqrt{(\Delta RA)^2 + (\Delta DEC)^2}$ from Sgr A*:
\begin{equation}
\Sigma(R) = A\left(\frac{R}{R_b}\right)^{-\Gamma}\left(1+(R/R_b)^{\delta}\right)^{(\Gamma-\beta)/\delta},
\label{eqn:profile}
\end{equation}
where A is the normalization, $R_b$ is the break radius, $\Gamma$ is the inner power-law index, $\beta$ is the outer power-law index, and $\delta$ is the sharpness of the transition between the two power-laws. Following the method in Appendix E of \citet{2013ApJ...764..154D}, we fit these parameters using Bayesian inference with the density profile as the likelihood for observing a given position. This allows us to infer the parameters of the fit given all the observed positional measurements. We use uniform priors for all the parameters and use a Metropolis-Hastings Markov Chain Monte Carlo (MCMC) method to sample the posterior distribution. Table \ref{tab:fit} summarizes our fit for the late-type and early-type stars. While we did not use binned points for the analysis, it is still a useful comparison to check for consistency with the fit. Figure \ref{fig:density} shows the radially binned surface density profile, which has been corrected for completeness as detailed in \citet{2013ApJ...764..154D} and \citet{2015ApJ...808..106S}. It also shows the projected surface density profile from the MCMC chains and the 1 $\sigma$ confidence interval in the model profile. The projected surface density of the late-type stars has central confidence interval for $\Gamma = -0.15_{-0.32}^{+0.24}$. The core radius, $R_b = 10.25_{-5.75}^{+3.31}$, is about 0.4 pc, beyond which the slope steepens to $\beta = 1.16_{-0.52}^{+0.61}$. Other observations in this region have also shown a lack of a stellar cusp \citep{2009A&A...499..483B,2010ApJ...708..834B}.

The flatness of the projected power-law density profile makes inference about the true 3D power-law slope difficult. Projection effects will flatten a wide range of slopes and even a negative power-law profile, or a ``hole'' in the density profile would generally appear flat in projection. In order to de-project the density profile, we use additional information from proper motion and radial velocity measurements. We incorporate this additional information using a spherically symmetric Jeans model of the cluster, including the effects of velocity anisotropy. The results of the Jeans modeling are detailed in \citet{2013ApJ...779L...6D}. From that analysis, the confidence interval for the three-dimensional slope, $\gamma = 0.05_{-0.60}^{+0.29}$. This value is significantly lower than the theoretical predictions of $\gamma = 3/2$ to $7/4$. An alternative method for modeling the three-dimensional density profile of the cluster uses acceleration measurements, which can directly constrain line-of-sight distances. This method offer tremendous potential and is presented in these proceedings by Chappell et al. 

If the density of red giants traces the underlying distribution of mass, then it would also imply a very flat density profile for the objects we cannot see, such as main sequence stars, white dwarfs, neutron stars, and stellar mass black holes. This would have strong implications for the dynamical friction time scale as the number of objects available to interact with is much smaller compared to a Bahcall-Wolf cusp. For example, the estimates for the merger rates of compact remnants with the supermassive black hole, a process which should produce strong gravitational waves \citep{2009MNRAS.395.2127O,2010ApJ...708L..42P,2012ApJ...757...27A}, would need to be modified, as they assume a cusp profile. Other dynamical calculations that depend on the underlying stellar density profile include the rate of two-body encounters and the resonant relaxation timescale \citep[e.g.][]{2005PhR...419...65A,2011ApJ...738...99M}. 

\begin{table}
\centering
\begin{tabular}{clccc}
\hline
Parameter & Description & Prior limits & Late-type (Old) & Early-type (Young)  \\
\hline
$R_b$ & break radius (arcsec) & [3, 15] & $10.25_{-5.75}^{+3.31}$ & $6.19_{-1.95}^{+2.26}$\\
$\Gamma$ & inner power-law index & [-1, 1] & $-0.15_{-0.32}^{+0.24}$ & $0.71_{-0.19}^{+0.14}$ \\
$\delta$ &  sharpness in transition & [0.5, 15] & $2.25_{-1.21}^{+3.68}$ & $5.46_{-3.84}^{+4.54}$ \\
$\beta$ & outer power-law index  & [0, 4] & $1.16_{-0.52}^{+0.61}$ & $2.89_{-0.76}^{+0.71}$ \\
$A$ & normalization (stars/sq. arcsec) & & $3.92_{-1.18}^{+4.43}$ & $0.89_{-0.29}^{+0.49}$ \\
\hline
\end{tabular}
\caption{Model Broken Power-law parameters, priors, and fits}
\label{tab:fit}

\end{table}

\section{The Paradox of Youth}
\label{sec:young}

The existence of very young stars in Galactic center has been termed the paradox of youth because the supermassive black hole should inhibit star formation. The tidal forces from the $4\times10^6$ M$_\odot$ black hole would prevent the type of molecular cloud seen in the solar neighborhood from collapsing to form stars. Two scenarios have been proposed to explain the existence of the young stars: (1) the stars are formed in-situ in a massive accretion disk \citep{2003ApJ...590L..33L,2009MNRAS.394..191H}, or (2) the young stars are deposited from an in-falling massive cluster originally formed further away \citep{2001ApJ...546L..39G,2006ApJ...650..901B}. These models make several predictions that can be tested observationally. An infalling cluster would likely leave a trail of stars during its inspiral into the Galactic center, resulting in a shallow surface density profile, falling as $R^{-0.75}$ \citep{2006ApJ...650..901B}. This profile should be shallower than if the young stars were formed in an accretion disk. Additionally, simulations of star formation in a massive accretion disk around Sgr A* suggest that the initial mass function of the young stars may top-heavy compared to the canonical Salpeter mass function \citep{2008Sci...321.1060B}. 

\begin{figure*}[tbh]
\center
\includegraphics[width=5.25in]{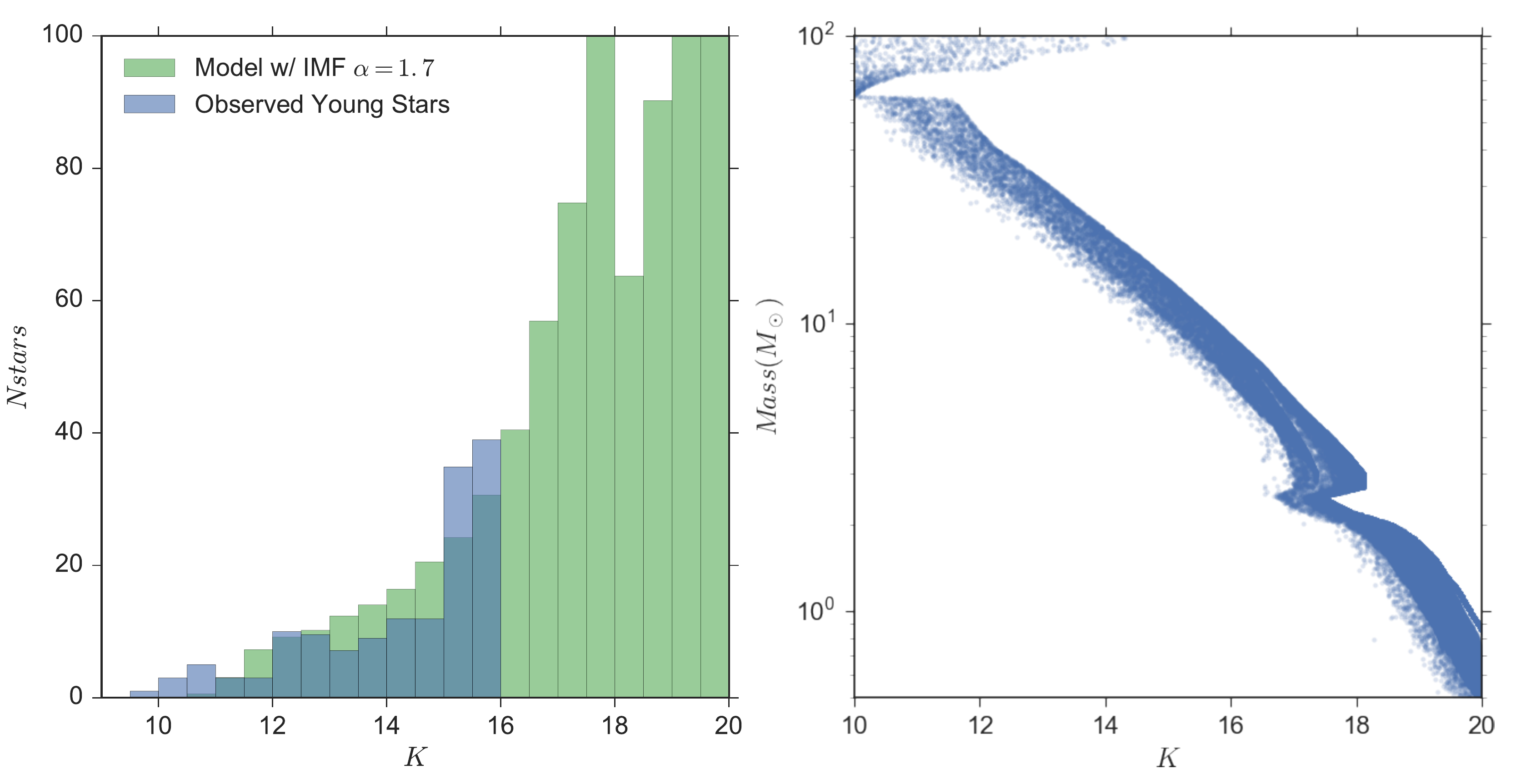}
\caption{\textbf{Left}: the observed completeness corrected luminosity function of young stars in the central 0.5 pc (blue) with the best fit stellar population synthesis model with a top-heavy mass function of $\alpha = 1.7$. \textbf{Right}: the relationship between the mass of a star and the observed K magnitude at the distance to the Galactic center and an extinction $A_K = 2.7$ mag. The relationship between luminosity and mass varies because some systems are in binaries.}
\label{fig:young_stars}
\end{figure*}

By spectroscopically separating the young stars from the old stars, it is possible to determine both the density profile and the luminosity function to test the two formation scenarios. The surface density profile of the young stars show that they are concentrated inside of $\sim 0.5$ pc, with a steep drop off beyond that radius. In fact, Gemini NIFS adaptive optics observations from  \citet{2015ApJ...808..106S} showed that the drop off is consistent with an edge to the young stellar cluster and almost no young stars are found outside of 0.5 pc (Figure \ref{fig:density}). The broken power-law fit has a break radius $R_b = 6.19_{-1.95}^{+2.26}$, or about 0.25 pc. Recent spectroscopic observations with KMOS on VLT and IRCS on Subaru have also shown a similar lack of young stars at large radii \citep{2015A&A...584A...2F,2016A&A...588A..49N}. The cut-off in the surface density profile supports the \textit{in-situ} star formation hypothesis, as no trail of stars is observed. The luminosity function of the young stars shows significantly fewer faint stars (i.e. less massive) than expected from a Saltpeter initial mass function. Using a stellar population synthesis model and accounting for binaries, \citet{2013ApJ...764..155L} found that the observed luminosity function corresponds to a mass function with a power-law slope of $\alpha = 1.7\pm0.2$ (compared to Salpeter of $\alpha = 2.35$, Figure \ref{fig:young_stars}). Observations from VLT show an even more top-heavy mass function \citep{2010ApJ...708..834B}. Having a top-heavy mass function likely means that the physical conditions for forming these stars are different than in the solar neighborhood; this lends additional support for star formation from a massive accretion disk. 

While the current favored scenario is that of in-situ star formation, there are a number of remaining issues both observationally and theoretically. Observationally, the spectroscopic measurements are largely limited only to stars brighter than about $K = 15.5-16$ mag, which corresponds to about 10 $M_\odot$ (Figure \ref{fig:young_stars}). A more robust measurement of the initial mass function requires better knowledge of the lower mass stars. There can also be significant degeneracies between star formation history, age, multiplicity, and initial mass function. Theoretically, it is difficult to explain why only about 20\% of the young stars are currently in a stellar disk around the black hole \citep{2014ApJ...783..131Y}. Star formation in a massive accretion disk should result in most of the stars being in the plane of the disk. Dynamical evolution is unlikely to eject many stars within the $4-6$ Myr age of the young star cluster. Perhaps not all of the stars were formed in the accretion disk, but some may have formed during the initial collapse of the molecular cloud that gave rise to the accretion disk. Additional simulations will be helpful to resolve these issues. 

\section{The Origin of the MW NSC}

There are two main theories for NSC formation: (1) they are the result of the infall of globular clusters that accumulate over time \citep[e.g.][]{2008ApJ...681.1136C,2001ApJ...552..572L},  and (2) they are formed in-situ from the infall of gas from the disk of the galaxy \citep[e.g.][]{2004ApJ...605L..13M}. Some combination of these two scenarios may also take place. Observationally constraining these scenarios is challenging. Estimates of the star formation history based on the luminous red giants suggest that most of the stars formed $> 1$ Gyr ago \citep{2007ApJ...669.1024M,2011ApJ...741..108P}, but it is difficult to determine whether they were formed at the Galactic center or elsewhere. A promising method to determine the origin of the cluster may be in the abundance patterns of the stars. If the stars are deposited from infalling globular clusters or dwarf galaxies, their abundance patterns should be  different than if they migrated in from the disk or formed from gas in the Milky Way disk. 

\begin{figure*}[tbh]
\center
\includegraphics[width=5.25in]{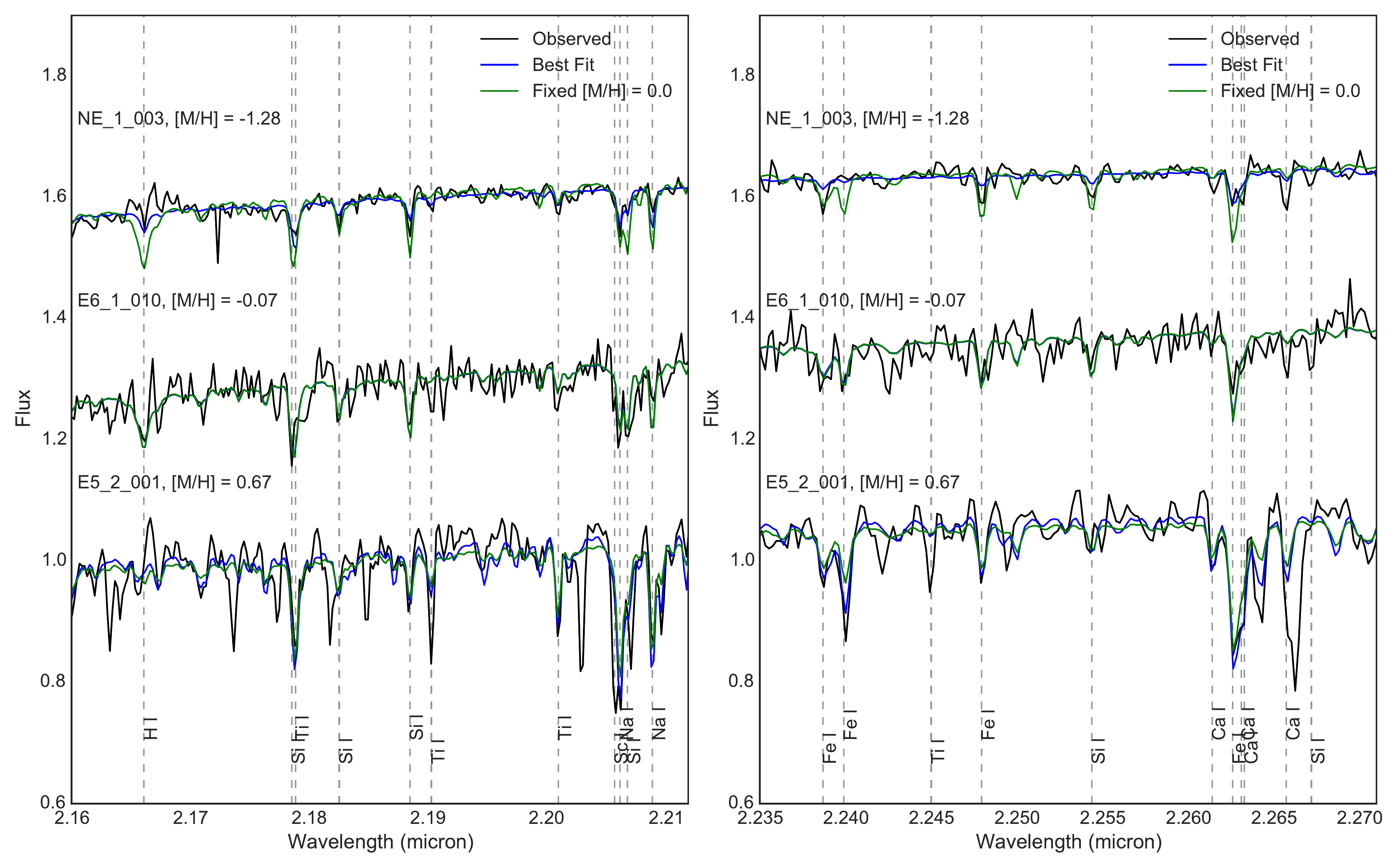}
\caption{Examples of spectra with a range of [M/H] along with their best fits (blue) compared to a fit with [M/H] fixed to 0.0 (green). Labeled are the best fit [M/H] values. The three stars show examples of metal-poor, solar metallicity, and super-solar metallicity stars. These two wavelength regions of K-band were chosen to illustrate the combination of temperature-sensitive lines such as H I compared to metallicity sensitive lines such as Fe I and Si I.}
\label{fig:spectra}
\end{figure*}

\begin{figure*}[tbh]
\center
\includegraphics[width=5.0in]{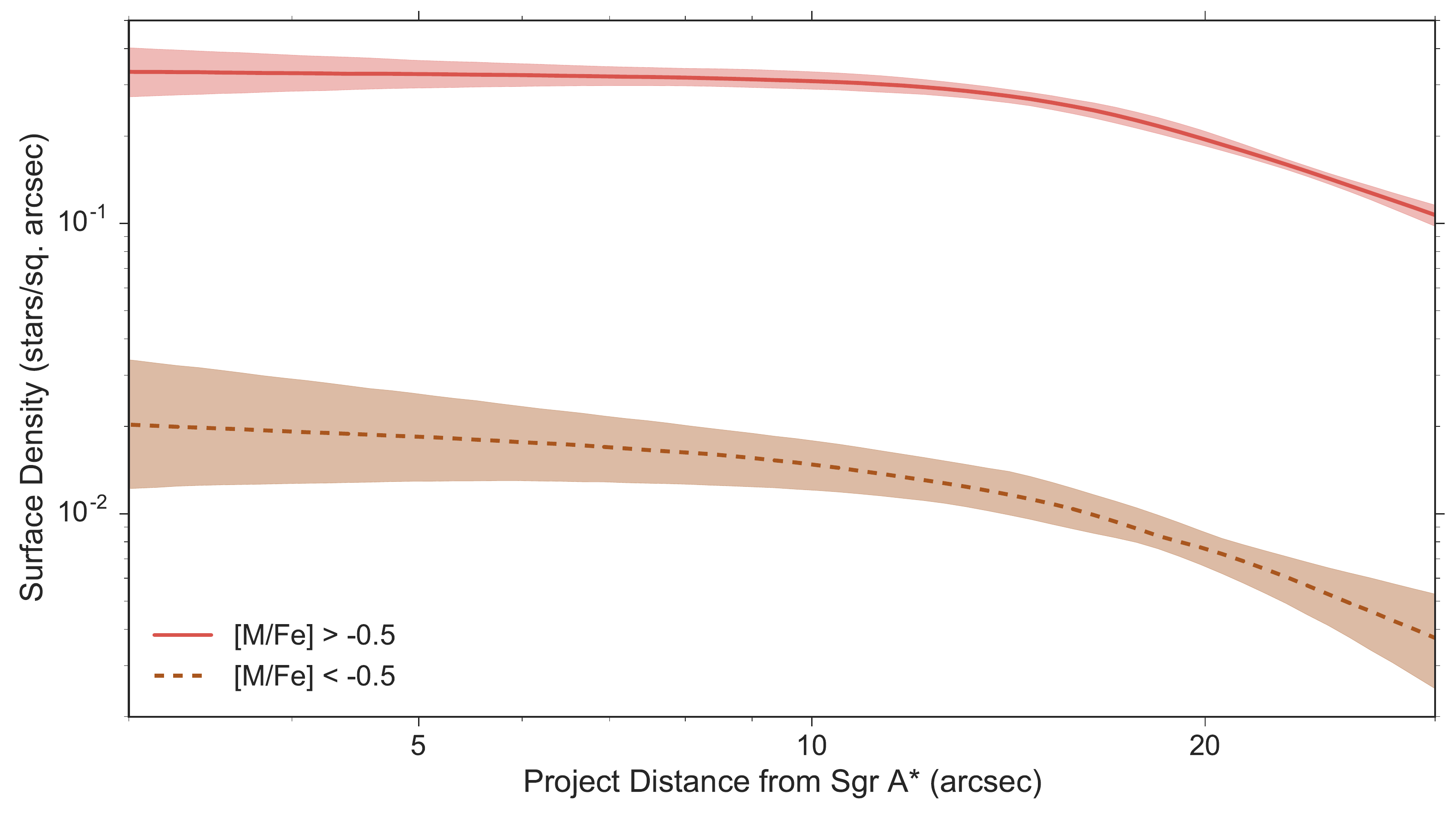}
\caption{The best fit surface density profile for high-metallicity ([M/Fe] $> -0.5$, solid red) and low-metallicity ([M/Fe] $< -0.5$, dashed brown) stars as a function of projected distance from the center of the cluster. The shaded regions are the 1 $\sigma$ model uncertainties a each given radius. Both density profiles are consistent within measurement uncertainties.}
\label{fig:metallicity_profile}
\end{figure*}

Using medium-resolution K-band spectra from NIFS with adaptive optics on Gemini North, \citet{2015ApJ...809..143D} measured the metallicity of about 80 stars in the central 1 pc of the MW NSC (Figure \ref{fig:spectra}). This greatly expanded on the number of stars with metallicities measured in this region. The observations showed that about 6\% of the stars have a scaled solar metallicity, [M/Fe] $< -0.5$ dex. This was the first time that low-metallicity stars were detected at the Galactic center. More recently, seeing-limited KMOS observations from VLT have also shown a similarly small fraction ($\sim5$\%) of low-metallicity stars \citep{2016arXiv161001623F}. While these observations are of brighter stars than the AO observations because of the confusion limit, they span a larger radial range, out to $\sim1.3$ pc in projected distance from the center of the cluster. Using this sample, we fit the surface density profile of the low metallicity compared to the high metallicity populations using Equation \ref{eqn:profile}. Within uncertainties, the surface density profile of the two population are consistent, including the slopes of the power-law and the break radius (Figure \ref{fig:metallicity_profile}). Table \ref{tab:metallicity} summarizes these fits. 

While low metallicity stars are consistent with that of globular clusters seen today \citep{1996AJ....112.1487H}, they represent only a small fraction of the cluster. The majority of the stars have greater than solar metallicity. This likely means that the cluster is not built up from globular clusters. High-resolution spectroscopic observations of these stars are important in order to obtain more accurate measurements of the metallicity and will allow measurements of individual elemental abundances \citep[e.g.][]{2015A&A...573A..14R,2015A&A...584A..45S}. This will allow a better determination of the cluster origin through comparisons of the Galactic center abundance patterns with large spectroscopic surveys such as APOGEE \citep{2014ApJ...796...38N}.

\begin{table}
\centering
\begin{tabular}{clccc}
\hline
Parameter & Description & Prior limits & [M/Fe] $> -0.5$ & [M/Fe] $< -0.5$  \\
\hline
$R_b$ & break radius (arcsec) & [3, 20] & $15.93_{-3.13}^{+2.47}$ & $16.10_{-5.49}^{+2.72}$\\
$\Gamma$ & inner power-law index & [-0.5, 1] & $0.04_{-0.22}^{+0.18}$ & $0.09_{-0.35}^{+0.42}$ \\
$\delta$ &  sharpness in transition & [1, 15] & $8.12_{-4.11}^{+4.49}$ & $8.69_{-4.68}^{+3.60}$ \\
$\beta$ & outer power-law index  & [-1, 4] & $1.65_{-0.37}^{+0.37}$&  $2.16_{-0.98}^{+1.19}$\\
$A$ & normalization (stars/sq. arcsec) & & $0.31_{-0.05}^{+0.07}$ & $0.01_{-0.001}^{+0.01}$ \\
\hline
\end{tabular}
\label{tab:metallicity}
\caption{Model broken power-law parameters, priors, and fits for metallicity sample}
\end{table}

\bibliography{/u/tdo/Documents/bibtex/prelim}

\end{document}